\documentclass[aps,prx,floatfix,reprint,twocolumn,showpacs,preprintnumbers,superscriptaddress,amsmath,amssymb]{revtex4-1}

\usepackage{graphicx}  % Include figure files
\usepackage{dcolumn}   % Align table columns on decimal point
\usepackage{bm}        % bold math
\usepackage{stmaryrd}  % for mapsfrom symbol
\usepackage{multirow}  % multirow tabular
\usepackage{color}     % color letters
\usepackage{amsmath,amssymb}
\usepackage{enumerate}
\usepackage{float}
\usepackage{mathtools}

\newcommand{\der}[3]{\frac{{\rm d}^{#1} #2}{{\rm d} #3^{#1}}}     % n-th derivation
\newcommand{\pder}[2]{\frac{\partial #1}{\partial #2}}            % 1st partial derivative
\newcommand{\fder}[2]{\frac{\delta #1}{\delta #2}}                % 1st functional derivative
\def   \blM   {{\bm M}}
\def   \blm   {{\hat{\bm m}}}
\def   \Aex   {{A_{\rm ex}}}
\def   \Ku    {{K_{u}}}

\def   \ez    {\hat{\bm e}_z}

\def   \tor   {{\bm \tau}}

\def   \bOm   {{\bm \Omega}}
\def  \gammag {{\gamma_{\rm g}}}

\def  \Ku     {K_{\rm u}}
\def  \Kp     {K_{\perp}}

\def   \dz      {\Delta{z}}
\def   \zc      {z_{\rm c}}

\def   \dtheta  {\delta{\theta}}
\def   \zex     {z_{\rm ex}}

\def   \Tuu   {T_{\uparrow\uparrow}}
\def   \Tud   {T_{\uparrow\downarrow}}
\def   \Tdu   {T_{\downarrow\uparrow}}
\def   \Tdd   {T_{\downarrow\downarrow}}

\begin{document}

\preprint{APS/123-QED}

\title{Domain wall dynamics due to femtosecond laser-induced superdiffusive spin transport}

\author{Pavel Bal\'a\v{z}} \email{balaz@karlov.mff.cuni.cz}
\affiliation{Charles University, Faculty of Mathematics and Physics, Department of Condensed Matter Physics, Ke Karlovu 5, CZ 121 16 Prague, Czech Republic}
\affiliation{IT4Innovations Center, VSB Technical University of Ostrava, 17.\ listopadu 15, CZ 708 33 Ostrava-Poruba, Czech Republic}
\author{Karel Carva}
\affiliation{Charles University, Faculty of Mathematics and Physics, Department of Condensed Matter Physics, Ke Karlovu 5, CZ 121 16 Prague, Czech Republic}
\author{Ulrike Ritzmann}
\affiliation{Department of Physics and Astronomy, Uppsala University, Box 516, SE-75120 Uppsala, Sweden}
\author{Pablo Maldonado}
\affiliation{Department of Physics and Astronomy, Uppsala University, Box 516, SE-75120 Uppsala, Sweden}
\author{Peter M.\ Oppeneer}
\affiliation{Department of Physics and Astronomy, Uppsala University, Box 516, SE-75120 Uppsala, Sweden}

\date{\today}

\begin{abstract}
Manipulation of magnetic domain walls via a helicity-independent laser pulse has recently been experimentally demonstrated and various physical mechanisms leading to domain wall dynamics have been discussed. Spin-dependent superdiffusive transport of hot electrons has been identified as one of the possible ways how to affect a magnetic domain wall.
Here, we develop a model based on superdiffusive spin-dependent transport to study the laser-induced transport of hot electrons through a smooth magnetic domain wall.
We show that the spin transfer between neighboring domains can enhance ultrafast demagnetization
in the domain wall. More importantly, our calculations reveal that when the laser pulse is properly focused on to the vicinity of the domain wall, it can excite sufficiently strong spin currents to generate a spin-transfer torque that can rapidly move the magnetic domain wall by several nanometers in several hundreds of femtoseconds, leading to a huge nonequilibrium domain wall velocity.
\end{abstract}


\maketitle

\section{Introduction}
\label{Sec:Intro}

Modern technologies such as magnetic memories or storage disks rely on the control and manipulation of magnetic bits. 
The ever-increasing demand for faster speed of magnetic recording continues to drive
the interest in finding improved ways to control the magnetization dynamics, either by using magnetic fields, spin-current torques~\cite{Slonczewski1996,Berger1996}, thermal gradients~\cite{Hatami2007,Moretti2017:PRB,HIN-11}, spin-orbit torques~\cite{Manchon2009}, or exchange-coupling torques \cite{Yang2015}. These advanced methods allow to construct a memory device based on manipulating domain walls~\cite{Parkin2008}.
The mutual interaction between light and magnetism, 
has gained much attention in the last twenty years initiated by the discovery  of Beaurepaire and coworkers~\cite{Beaurepaire1996} of ultrafast laser-induced demagnetization of Ni thin films. This phenomenon plays a dominant role, too, in the discussions on future magnetic devices. As a consequence, many related experimental studies have been carried out, leading to the discoveries of e.g.\ the optical spin-transfer torque~\cite{r_12_Nemec_opticalSTT,Janda2017} and the optical spin-orbit torque~\cite{r_13_Tesarova_Nemec_Observ_SOT}.

From a theoretical point of view, different models and mechanisms were proposed to explain the ultrafast laser-induced demagnetization 
\cite{Beaurepaire1996,Koopmans2010,Mueller2011,Carpene2008,Krauss2009,Roth2012,Carva2011PRL,Battiato2010,Battiato2012}. From those, it is important to emphasize two, namely, ultrafast demagnetization due to spin-flip relaxation~\cite{Beaurepaire1996,Koopmans2010,Mueller2011,Carpene2008,Krauss2009,Roth2012,Carva2011PRL} and non-local superdiffusive spin transport ~\cite{Battiato2010,Battiato2012}. More specifically, within the latter the model introduced by  Battiato, Carva, and Oppeneer~\cite{Battiato2010,Battiato2012} ultrafast demagnetization could be shown without any additional assumptions for a spin-flip scattering mechanism. Importantly, the non-local character of this mechanism invoking spatial spin transport  suggests that laser excitation can also be used to manipulate magnetic structures in magnetically non-homogeneous systems \cite{Rudolf2012,Schellekens2014,Choi2014,Razdolski2017}. For instance, it has been shown experimentally that the interplay between magnetic textures and ultrafast demagnetization can affect the domain structure in thin Co/Pt films on the sub-picosecond timescale~\cite{Pfau2012}. Additionally, it has been suggested that transfer of hot electrons flowing between different magnetic domains can accelerate the demagnetization process~\cite{r_12_Vodung_UltraDemagMultilay_SpinTransp}. On the one hand, experimental observations on various samples indicate that this effect might be limited in some materials~\cite{Moisan2014}.
On the other hand, more recently  laser-controlled manipulation of domain walls  has been
demonstrated in Co/Pt thin films~\cite{Quessab2018:PRB}, in Co/Cu/Ni trilayer films \cite{Sandig2016}, and in Co/Fe$_{75}$Gd$_{25}$ bilayer films \cite{Shokr2019}, as well as helicity-dependent domain wall motion \cite{Janda2017},  formation of vortices~\cite{Eggebrecht2017_LI_Textures},
and combination of  laser-driven domain wall  motion with spin-Hall effect~\cite{Lalieu2018:arXiv},
and with currents~\cite{Mangin2018:arXiv}.

In this paper, we study how superdiffusive spin currents can influence the magnetic texture, specifically,
 a magnetic domain wall.
For simplicity we assume a one dimensional model of an isolated magnetic domain wall separating
two domains with opposite orientation of magnetization.
Initially, we study demagnetization in a narrow magnetic domain wall to compare our results with
previous studies~\cite{Pfau2012}.
Subsequently, we extend the study to wider and more realistic magnetic domain walls taking into account their detailed structure.
To this end we generalized the model of superdiffusive spin-dependent transport for the 
study of noncollinear magnetic configurations. Our work reveals how a domain wall influences the flux of electrons along the sample and the demagnetization process. In more detail, we calculate the nonhomogeneous spin transfer torque acting on the domain wall
and, consequently, study the dynamics of the magnetic moments.
We show that when a femtosecond laser pulse is focused properly, it can
trigger domain wall motion and shift the domain wall center with a very high out-of-equilibrium velocity of about $10^4$ m/s.

The paper is organized as follows. In the following Section we provide a short introduction to the superdiffusive spin-dependent transport model of hot electrons.
In Section~\ref{Sec:Model} we describe the generalization of the model for noncollinear magnetic textures.
In Sec.~\ref{Sec:NarrowDW} we study the influence of the domain wall on ultrafast demagnetization.
Second, in Sec.~\ref{Sec:WideDW} we study ultrafast demagnetization and spin-transfer torque generation
in wide magnetic domain walls. The laser-induced magnetization dynamics is studied as well.
Finally, we discuss the consequences of our results in Sec.~\ref{Sec:Discussion}.

\section{Methods}

The superdiffusive spin-dependent transport model~\cite{Battiato2010,Battiato2012} is based on the distinct transport behavior of minority and majority electrons due to their different velocities and lifetimes.
The starting point of the superdiffusive spin-dependent transport model of hot electrons~\cite{Battiato2010,Battiato2012}
is the excitation of localized electrons above the Fermi level (usually from $d$ to $s$ band) induced by a femtosecond laser pulse.
Because of the higher electron velocities the excited hot electrons are treated
as itinerant particles moving along the sample~\cite{Battiato2014_JAP}.
Motion of the itinerant electrons is described by a transport equation taking into account
electrons spins, $\sigma \in \{\uparrow, \downarrow\}$, and energies, $\epsilon$.
The  electrons velocities, $v_\sigma(\epsilon)$, 
and lifetimes, $\tau_\sigma(\epsilon)$ depend on these quantities.
Due to the difference of $v_\uparrow$ ($\tau_\uparrow$) and $v_\downarrow$ ($\tau_\downarrow$)
in a magnetic material,  the current of flowing electrons gets polarized.
In case of multilayers, spin filtering via multiple spin-dependent reflections at the
interfaces contributes to the spin current polarization~\cite{Battiato2014_JAP}. 
As a result, loss of the magnetic momentum carried away by the spin currents
is observed as a local demagnetization of the magnetic material illuminated by the laser pulse.
Due to high electron velocities, this demagnetization process happens on a femtosecond time scale.
Note, this model assumes purely non-thermal laser excitation of electrons without any effects of temperature~\cite{Moreno2016}.
Importantly, the spin transport of the hot electrons is neither purely ballistic nor diffusive.
Its transport character is changing in time from initially ballistic motion towards diffusive motion via the superdiffusive regime, which
takes between $500\, {\rm fs}$ up to $1\,{\rm ps}$ depending on 
the laser pulse and material properties~\cite{Battiato2012}.

The model has been supported by a number of experimental observations~\cite{Schellekens2014,Choi2014,Bergeard2016,Eschenlohr2017,Lalieu2017,Hofherr2017,Malinowski2018}.
% Experiments~\cite{Lalieu2017,Razdolski2017} and numerical simulations~\cite{Ulrichs2018} 
% have revealed that spin currents induced by a laser pulse can trigger 
% terahertz spin waves in noncollinear spin valves.
Alternatively, the theory of spin dependent transport of hot electrons has been formulated 
in the framework of Boltzmann transport equation~\cite{Nenno2016}
showing that the energy dependence of the injected hot electrons plays a crucial role 
for the spin transport.
% A different theoretical model, based of spin-Vlasov equation,~\cite{Hurst2014:EPJD}
% imply that even laser pulse of a modest amplitude 
% can generate spin currents in ferromagnet in thin magnetic films.~\cite{Hurst2018:PRB}

So far, the superdiffusive transport model has been mostly used to explain ultrafast demagnetization processes,
especially in single magnetic layers and collinear magnetic multilayers.
However, in case of noncollinear magnetic configuration the spin currents generated by the
superdiffusive spin-dependent transport of hot electrons can induce 
spin-transfer torques~\cite{Slonczewski1996,Berger1996,r_02_Slon_CurrTor,r_02_StilesAnatomy,r_08_Ralph_Stiles_STTreview,r_93_vf,Barnas2005}
acting on the magnetic moments and, consequently, magnetization dynamics~\cite{Berkov_Miltat_2008}.
Such a noncollinear magnetic configuration can be achieved in magnetic multilayers or
in magnetic films or wires featuring magnetic textures like domain walls, magnetic bubbles or skyrmions.

Recently, we developed an effective model for the spin-transfer toque induced by hot electrons in noncollinear spin valves~\cite{Balaz2018:JPCM}.
It has been shown that, in accord with experimental observations \cite{Rudolf2012,Schellekens2014,Choi2014,Razdolski2017}, excitation of hot electrons in one magnetic layer
can lead to a fast spin-transfer torque and small angle magnetization precessions in the second magnetic layer  even though
both magnetic layers are separated and magnetically decoupled by a nonmagnetic one. 
Here, we adopt a different approach where we take into account spin rotation between neighboring magnetic moments in a
magnetic domain wall. The spin-dependent transport properties inside the domain wall are then included in the transmissions and reflections between discretization cells with uniform magnetizations as described below in Sec.~\ref{SSec:transport}.

\section{Theoretical model}
\label{Sec:Model}

%%%%%%%%%%%%%%%%%%%%%%%%%%%%%
\begin{figure}[!htp]
 \centering
 \includegraphics[width=.8\columnwidth]{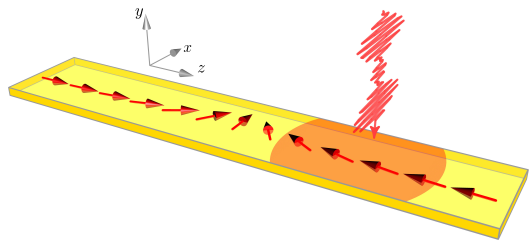}
 \caption{(Color online) Sketch of a head-to-head magnetic domain wall and the used coordination system.
          The wavevector  of the incident laser beam is aligned with the $y$ axis.}
 \label{Fig:DWscheme}
\end{figure}
%%%%%%%%%%%%%%%%%%%%%%%%%%%%%
We start with the investigation of laser-induced magnetization dynamics of a 1-dimension wire 
of length $L = N \dz$, where $\dz$ is a spatial discretization length and $N$ is the number of cells in the simulation, which simulates a domain wall. The domain wall is located in the middle of the wire, where $z = \zc$. We illustrate the domain wall schematically in Figure \ref{Fig:DWscheme}.

We assumed that the magnetization is varying along the $z$ axis.
In equilibrium the magnetization vector is given by $\blM(z) = M_0\, \blm(z)$, where $M_0$ is the saturated magnetization
and $\blm(z)$ is a unit vector
\begin{equation}
  \blm(z) = (\cos\phi \sin\theta(z), \sin\phi \sin\theta(z), \cos\theta(z))\,,
\label{Eq:magndir}
\end{equation}
where $\phi$ is  constant with respect to $z$, and 
\begin{equation}
  \theta(z) = 2 \arctan \left[ \exp \left( \frac{z - z_0}{\Delta} \right) \right] , \\
\label{Eq:theta_z}
\end{equation}
with $z_0$ the position of the domain wall center.
$\Delta$ is the domain wall width given by 
\begin{equation}
  \Delta = \sqrt{\Aex / \Ku}\,,
\label{Eq:DW_width}
\end{equation}
with $\Aex$ being the exchange stiffness and $\Ku$ is the uniaxial anisotropy constant~\cite{SchryerWalker1974,Tatara:SciRep_2008}.
Eq.~(\ref{Eq:theta_z}) describes a head-to-head magnetic domain wall separating the left magnetic domain
with magnetization $\blm_{\rm L} = \ez \equiv (0,0,1)$ from the right one 
$\blm_{\rm R} = -\ez$.

For our specific 1-dimensional wire, we assume $\phi = \pi/2$ (the magnetization is in the layer's plane).
Thus the magnetization in the $i$-{\em th} cell reads $\blm_i = (\sin\theta_i, 0, \cos\theta_i)$; see Fig.~\ref{Fig:DWscheme}, and the magnetization direction $\theta_i \equiv \theta(z_i)$ becomes
\begin{equation}
  \theta_i = 2 \arctan \left[ \exp \left( \frac{z_i - \zc}{\Delta} \right) \right]\,, 
  \quad i = 1, 2, \dots, N ,\\
\label{Eq:thetai}
\end{equation}
where $z_i$ is the position of the $i$-th cell.

\subsection{Electronic transport}
\label{SSec:transport}

In each discretization cell, the quantization axis is aligned along the local magnetization, $\blm_i$.
We assumed that the electron spins are always parallel to the direction of the local quantization axis.
Thus, only the longitudinal part of the spin current from the neighboring cells enters the $i$-th cell.
The transverse part gives rise to the spin torque, see below.
The change of the quantization axis, encountered by the electrons moving along the wire,
is described by transmission matrices defined at the interfaces between the cells allowing reversal of electrons' spins.
The interface transmission matrix between $i$-th and $(i+1)$-th cell is given by
\begin{equation}
  {\bm T}^{(i)} = 
  \begin{pmatrix}
    \Tuu^{(i)} & \Tud^{(i)} \\
    \Tdu^{(i)} & \Tdd^{(i)}
  \end{pmatrix} ,
\label{Eq:TransMatrix}
\end{equation}
where $i = 1, 2, \dots, N-1$, and
\begin{subequations}
  \begin{align}
    \Tuu^{(i)} = \Tdd^{(i)} &= \cos^2(\dtheta^{(i)}/2)\,, \\
    \Tud^{(i)} = \Tdu^{(i)}   &= \sin^2(\dtheta^{(i)}/2)\,,
  \end{align}
\label{Eqs:transmissions}
\end{subequations}
with
$\dtheta^{(i)} = \theta_{i+1} - \theta_i$.

The transmission matrix (\ref{Eq:TransMatrix}) applies for electrons moving in both directions.
Therefore, when the spin current in the $i$-th cell moving from the left to the right reads
\begin{equation}
  \overrightarrow{{\bm\jmath}_i} = 
  \begin{pmatrix} 
    \overrightarrow{\jmath_{\uparrow i}}\\ 
    \overrightarrow{\jmath_{\downarrow i}}
  \end{pmatrix} ,
\end{equation}
 the spin current in the $(i+1)$-th cell reads
\begin{equation}
  \overrightarrow{{\bm\jmath}_{i+1}} = {\bm T}^{(i)} \cdot \overrightarrow{{\bm\jmath}_i}\,.
\label{Eq:curr_right}
\end{equation}
Similarly, for the current moving from the right to the left we can write
\begin{equation}
  \overleftarrow{{\bm\jmath}_{i}} = {\bm T}^{(i)} \cdot \overleftarrow{{\bm\jmath}_{i+1}}\,.
\label{Eq:curr_left}
\end{equation}

Note that the currents in Eqs.~(\ref{Eq:curr_right}) and (\ref{Eq:curr_left}) are taken at the same energy
level $\epsilon_j$. 
We assume that electrons flowing through an interface do not change their energy.

\subsection{Spin-transfer torque}

The spin-transfer torque acting on magnetization $\blm_i$ in the $i$-th cell 
is proportional to the transverse (with respect to $\blm_i$) spin current flowing into the $i$-th cell from $(i-1)$-th and $(i+1)$-th ones.

We first define the spin current from electrons moving from the right to the left in the $i$-th cell.
Since the spin current is defined locally we shall distinguish between spin currents flowing
through the left interface of the $i$-th cell, labeled by $\xi = (-)$, and through the
right interface of the $i$-th cell, labeled by $\xi = (+)$.
Generally, we define the spin current of right-moving electrons in the $i$-th cell as
\begin{equation}
  \overrightarrow{\jmath_{s\,(i)\, \xi}} = \frac{\hbar}{2\, e} \sum_{j=1}^{N_\epsilon} 
  \left[ \overrightarrow{\jmath_{\uparrow\,(i)\, \xi}}(\epsilon_j) - \overrightarrow{\jmath_{\downarrow\,(i)\, \xi}}(\epsilon_j) \right]\,,
\label{Eq:sc_right}
\end{equation}
where $N_\epsilon$ is the number of energy levels assumed in the calculations.
Similarly, we can define spin currents flowing from the right to the left
\begin{equation}
  \overleftarrow{\jmath_{s\,(i)\, \xi}} = \frac{\hbar}{2\, e} \sum_{j=1}^{N_\epsilon} 
  \left[ \overleftarrow{\jmath_{\uparrow\,(i)\, \xi}}(\epsilon_j) - \overleftarrow{\jmath_{\downarrow\,(i)\, \xi}}(\epsilon_j) \right]\,.
\label{Eq:sc_left}
\end{equation}

It is important to note, by definition, currents of electrons moving from the left to the right, 
$\displaystyle \overrightarrow{\jmath_{\sigma\, (i)\, \xi}}(\epsilon_j)$, have positive sign while
the currents moving from the right to the left, 
$\displaystyle \overleftarrow{\jmath_{\sigma\, (i)\, \xi}}(\epsilon_j)$, have negative sign.

Now, let us define the spin transfer torque acting on magnetization in $i$-th cell. 
From the right moving electrons we have the contribution
\begin{equation}
  \overrightarrow{\tor_{i}} = \overrightarrow{\jmath_{s\,(i-1)\,(+)}}\, \blm_{i-1} - \overrightarrow{\jmath_{s\,(i)\,(-)}}\, \blm_i\,,
\label{Eq:def_tor_right}
\end{equation}
while from the left moving electrons we obtain
\begin{equation}
  \overleftarrow{\tor_{i}} = \overleftarrow{\jmath_{s\, (i+1)\, (-)}}\, \blm_{i+1} - \overleftarrow{\jmath_{s\, (i)\, (+)}}\, \blm_i\,.
\label{Eq:def_tor_left}
\end{equation}
The total spin torque action on $i$-th magnetization is given by
\begin{equation}
  \tor_i = \overrightarrow{\tor_{i}} + \overleftarrow{\tor_{i}}\,,
\end{equation}
which gives
\begin{equation}
  \begin{split}
  \tor_{i} = &\,\overrightarrow{\jmath_{s\,(i-1)\, (+)}}\, \blm_{i-1} 
    - \left( \overrightarrow{\jmath_{s\,(i)\, (-)}} + \overleftarrow{\jmath_{s\, (i)\, (+)}} \right)\, \blm_i\; + \\
    & \overleftarrow{\jmath_{s\, (i+1)\, (-)}}\, \blm_{i+1}\,.
  \end{split}
\label{Eq:def_tor}
\end{equation}

For the sake of clarity, in the definitions (\ref{Eq:sc_right})--(\ref{Eq:def_tor}) we omitted the time dependence
of the electron currents. The magnetizations directions, $\blm_i$, are assumed to be constant in the transport calculations.

\subsection{Magnetization dynamics}

To study the magnetization dynamics of the magnetic cells we use the Landau-Lifshitz-Gilbert equation (LLG), which for the $i$-th magnetic moment in the wire reads
\begin{equation}
    \der{}{\blm_i}{t} - \alpha\, \blm_i \times \der{}{\blm_i}{t} = \bOm_i\,,
\label{Eq:LLG}
\end{equation}
where $\gamma = |\gammag| > 0$ is the gyromagnetic ratio, $t$ is time, and
$\alpha$ is the Gilbert damping parameter, and $\bOm_i$ is the overall torque acting on the $i$-th local magnetization
defined as
\begin{equation}
  \bOm_i = -\mu_0\, \gamma\, \blm_i \times {\bm H}_{\rm eff}^{(i)} 
  + \frac{1}{\mu_0\, M_i^2\, V_{\rm cell}}\, \tor_i\,,
\label{Eq:torque}
\end{equation}
which consists of a part induced by the effective magnetic field, ${\bm H}_{\rm eff}^{(i)}$, and the spin-transfer torque term containing $\tor_i$, where
$M_i$ is the magnitude of magnetization in the $i$-th computational cell, and $V_{\rm cell}$ is the cell volume.
Here, it is important to mention that the magnitude $M_i$ depends on the superdiffusive transport and changes in time~\cite{Battiato2010}.
Equation (\ref{Eq:torque}) shows that spin-transfer torque is stronger when acting on a localized magnetic moment.

Generally, the effective magnetic field is defined as a functional derivative of total volume energy density, $w$,
\begin{equation}
  {\bm H}_{\rm eff}^{(i)} = \frac{1}{\mu_0\,M_0} \fder{w(z_i)}{\blm_i}\,,
\label{Eq:Heff}
\end{equation}
where $\mu_0$ is vacuum permeability, and $M_0$ is the equilibrium value of the saturated magnetization,
and $z_i$ is the position of the $i$-th cell.
The  energy density of the total volume reads
\begin{equation}
  \begin{split}
    w(z) =\; &\Aex \, \left( \pder{\theta(z)}{z} \right)^2 + 
           \Ku\, \cos^2\!\theta(z) + \\
           &\Kp\, \left(\cos\!\phi\, \sin\!\theta(z) \right)^2 ,
  \end{split}
\label{Eq:wDW}
\end{equation}
where $\Ku$ is the uniaxial anisotropy constant and $\Kp$ is the perpendicular out-of-plane anisotropy constant.
The spin-transfer torque, $\tor_i$, is given by Eq.~(\ref{Eq:def_tor}).

%%%%%%%%%%%%%%%%%%%%%%%%%%%%%%%%%%%%%%%%%%%%%%%%%%%%%%%%%%%%%%%%%%%%%%%%%%%%%%%%%%%%%%%%%%%%%%%%%%%

\section{Narrow magnetic domain wall}
\label{Sec:NarrowDW}

We apply the above described model to study the spin-dependent transport in a simplified system of a narrow magnetic domain wall to investigate the effects of a femtosecond laser pulse on magnetic domains walls formed typically in multilayers like Co/Pt or Co/Pd having a strong out-of-plane magnetic 
anisotropy~\cite{Pfau2012,r_12_Vodung_UltraDemagMultilay_SpinTransp,Moisan2014}.

In this case we focus on the effects of the superdiffusive transfer on the domain wall structure.
Thus we use first a simplified model of a sharp domain wall, where the magnetization direction is changed abruptly by $180$ degrees, which reproduces the experimental conditions. 
Hot electrons passing the domain wall move from being in the majority to being in the minority spin channel and vice versa.
As a result, spins accumulate in the vicinity of the domain wall
causing a change of the domain wall profile~\cite{Dugaev2002:PRB}.
No spin flips or reflections were assumed at the domain wall.

\begin{figure}[htp!]
 \centering
 \includegraphics[width=.9\columnwidth]{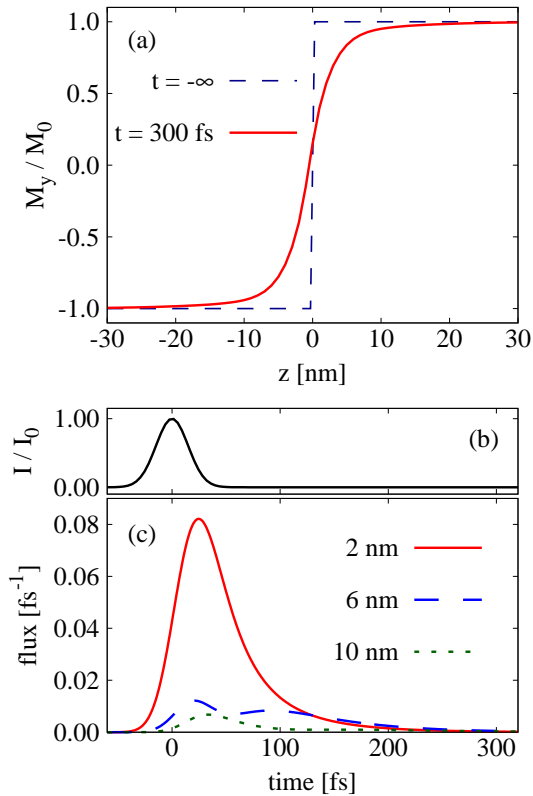}
 \caption{(Color online) Superdiffusive transport through a narrow magnetic domain wall.
          (a) Dashed (blue) line shows out-of-plane component of magnetization in the vicinity of a narrow magnetic domain wall in the equilibrium ($t = -\infty$).
          Solid (red) line shows the out-of-plane magnetic component $300\, {\rm fs}$ after the maximum of the laser pulse.
          (b) Time variation of Gaussian laser pulse intensity with peak at time $t = 0$ and FWHM $35\, {\rm fs}$.
          (c) Corresponding time variation of the spin fluxes calculated for different distances ($2$, $6$, and $10\, {\rm nm}$) from the domain wall.}
 \label{Fig:dw_narrow}
\end{figure}

In our calculations we assumed a sample as long as $100\, {\rm nm}$ with spatial discretization
$\Delta{z} = 1\, {\rm nm}$. Moreover, we assumed time discretization $\Delta{t} = 1\, {\rm fs}$.
For simplicity, we used the electron velocities and lifetimes of hot electrons
calculated for Fe~\cite{r_09_Zhuko_GW_SpinLifetimeSO}.
For the electronic transport we assumed $N_\epsilon = 12$ energy levels above the Fermi energy.
The difference between the subsequent energy levels was $\Delta\epsilon = 0.125\, {\rm eV}$,
which allows us to cover an energy range up to $1.5\, {\rm eV}$.
We assumed that the whole sample was homogeneously excited by the laser pulse having a Gaussian shape
with maximum in $t = 0$ and full width in half maximum $t_{\rm p} = 35\, {\rm fs}$; see Fig.~\ref{Fig:dw_narrow}(b).
Altogether, this pulse excites $0.2$ electrons at each energy/spin level, which corresponds to a laser fluence $F =  2.6$ mJ/cm$^2$.

In Fig.~\ref{Fig:dw_narrow}(a) we show the effect of the femtosecond spin pulse on the domain wall by locating the sharp step domain wall at $z=0$.
The dashed (blue) line shows the equilibrium distribution of the out-of-plane magnetization component along the sample normalized to 
the equilibrium saturated magnetization $M_0$.
In turn, the solid (red) line describes the out-of-plane magnetization $300\, {\rm fs}$ after the laser pulse intensity reached its maximum.
The boundary between the domains has become smeared due to ultrafast demagnetization caused by the superdiffusive spin transport
between the neighboring magnetic domains.
The hot electrons excited by the laser pulse move along the sample and carry the angular momentum.
Because the velocities of electrons in spin-up and spin-down channels differ, spin accumulation builds up in the vicinity
of the domain wall.
As a result, the domain wall profile becomes smeared. 
Fig.~\ref{Fig:dw_narrow}(a) shows that the spin accumulation decays as a function of the distance from the domain wall center
and reaches up to about $20\, {\rm nm}$.
This result is in a good agreement with experimental observations and Monte Carlo simulations reported by Pfau {\em et al}.~\cite{Pfau2012}.
%%%%%

The time scale of the ultrafast demagnetization is shown in Figs.~\ref{Fig:dw_narrow}(c),
which illustrates the time variation of the total spin flux taken at different distances from the domain wall.
The amplitude of the spin flux strongly decreases with the distance from the domain wall.
Especially, temporal dependences of more remote fluxes exhibit more than one peak.
This is a result of the secondary electrons which are generated by the avalanches during the scattering of the hot electrons.
Importantly, the out-of-equilibrium spin fluxes become zero after about $300\, {\rm fs}$ from the laser pulse, indicating the time scale of the ultrafast demagnetization.

%%%%%%%%%%%%%%%%%%%%%%%%%%%%%%%%%%%%%%%%%%%%%%%%%%%%%%%%%%%%%%%%%%%%%%%%%%%%%%%%%%%%%%%%%%%%%%%%%%%

\section{Wide magnetic domain wall}
\label{Sec:WideDW}

In the following, we present our numerical results for a wider domain wall with a detailed structure of the domain wall profile
as described by Eq.~(\ref{Eq:theta_z}).
In our calculations we assumed a wire as long as $200\, {\rm nm}$.
With spatial discretization $\Delta{z} = 1\, {\rm nm}$ we used $N = 200$ computational cells.
The excited electrons can occupy $N_\epsilon = 12$ energy levels above the Fermi energy
with energy discretization $\Delta\epsilon = 0.125\, {\rm eV}$.
The energy and spin-resolved electron velocities and lifetimes used in our calculations
correspond to those of iron  as calculated by \textit{ab initio} methods~\cite{r_09_Zhuko_GW_SpinLifetimeSO,r_08_Zhuko_GW_SpinRelaxExc_SO,Battiato2012}.

Moreover, in all our simulations we assume a Gaussian laser pulse of length $t_{\rm p} = 35\, {\rm fs}$.
During its duration, this pulse excites the same number of electrons on each energy and spin level.
When considering the laser fluence 
$F \simeq 12.8\, {\rm mJ}/{\rm cm}^2$
we obtain $1$ excited electron on each energy/spin level in each 1-nm wide discretization cell. 

\subsection{Ultrafast demagnetization}

\begin{figure}[!htp]
 \centering
 \includegraphics[width=.9\columnwidth]{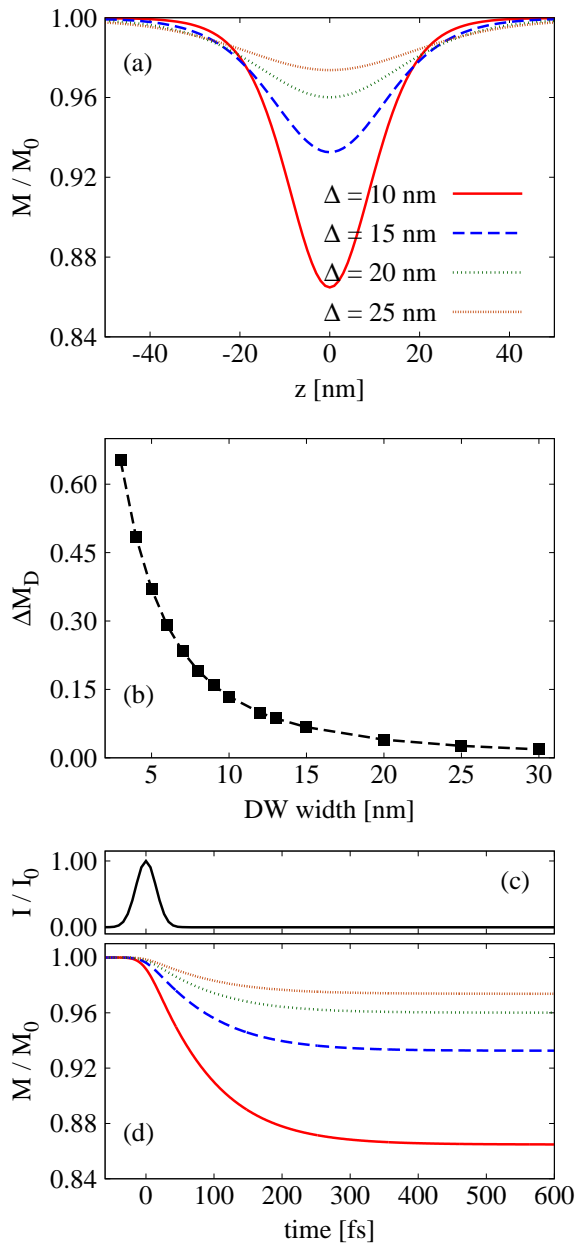}
 \caption{(Color online) Laser-induced demagnetization in the vicinity of the domain wall calculated for various DW widths.
          (a) Spatial dependence of magnetization taken $1\, {\rm ps}$ after the pulse.
          (b) Demagnetization in the DW center, $z = 0$, taken $1\, {\rm ps}$ after the pulse as a function of the DW width.
          (c) Time variation of Gaussian laser pulse intensity with peak at time $t = 0$ and FWHM $35\, {\rm fs}$.
          (d) Time-dependence of magnetization in the DW center after the laser pulse, for various DW widths. The lines correspond to those shown in (a).}
 \label{Fig:demag}
\end{figure}

First, we focus on the ultrafast demagnetization induced by the laser pulse.
Here, we assume that the whole sample is excited homogeneously. 
Therefore, the same number of electrons is excited by the laser pulse in each computational cell.
This situation corresponds to a case where the laser spot exceeds the size 
of the computational length of the wire, $L$.
In our analysis, we focus on the central part of the wire, which is far from its boundaries.
In the case of a uniformly magnetized sample,  the same amount of electrons is flowing in both directions in the center of the wire.
Thus, no demagnetization in the here-considered 1D transport model can be obtained.
Oppositely, when a domain wall of width $\Delta$ is located in the center of the wire,
electrons flowing through the domain wall from the left to the right are polarized in the left magnetic domain, while the electrons moving in the
opposite direction are polarized in the right magnetic domain.
Thus the left and right spin fluxes have opposite polarizations.
As a result, partial demagnetization at the position of the domain wall (DW) and its vicinity can be expected.
This mechanism is the same as the one described above for the narrow DW.
Fig.~\ref{Fig:demag}(a) shows the magnetization profile as function of the distance with respect to the central part of the wire taken $1\, {\rm ps}$ after the laser pulse
for various DW widths; $M_{\rm D}(z) = M(t = 1\, {\rm ps}, z)$.
%%%
%%%
We observe that the maximum demagnetization, 
$\Delta{M}_{\rm D} = 1 - M_{\rm D} / M_0$, is observed in the center of the domain wall and
decreases towards zero deeper in the domains.
Since the electronic transport is limited by the lifetimes of $s$-electrons, the
demagnetization becomes less pronounced for wider domain walls.
Fig.~\ref{Fig:demag}(b) shows how the maximum demagnetization in the DW center, $z = 0$,
changes as a function of the DW width.
This reflects the fact that the demagnetization is governed by the spin transport
between magnetic domains with opposite magnetizations.

Additionally, we focus on the time dependence of the demagnetization process.
Fig.~\ref{Fig:demag}(b) shows the time dependence of te magnetization in the DW center, $z = 0$.
In this point we define the demagnetization time $\tau_{\rm D}$ as a time when the
local magnetization $M(t)$ reaches $M(\tau_{\rm D}) = (1 - e^{-1})\, (M_0 - M_{\rm D})$. 
This demagnetization time has been found to be virtually the same for all the
studied DW thicknesses, $\tau_{\rm D} \simeq 145\, {\rm fs}$.

It is important to note that in a realistic system, when excited electrons moves in all three directions,
demagnetization due to inter-domain spin transport will appear as an effect additional 
to the demagnetization, which might be observed also in samples with uniform magnetization.

\subsection{Spin-transfer torque}

When a magnetic domain wall is located in the middle of the sample, which is uniformly excited by a laser pulse,
the equal flow of electrons in both directions leads to zero total spin-transfer torque acting on the domain wall.
Hence, no net domain wall motion can be expected due to a laser excitation in a symmetric system.
To obtain a nonzero spin-transfer torque, one needs to create asymmetry in the left and right electron fluxes.
This can be accomplished by a number of different ways using electric or thermal gradients, 
employing different materials or by changing magnetic topology in the sample creating additional magnetic textures.
Most of these methods, however, exert an additional torque on the magnetic domain wall.
Therefore, here we study a simplified model, by assuming that the hot electrons are excited by the laser pulse only in a certain restricted region of
the sample of width $l_{\rm ex}$. We also assume that electrons are excited homogeneously by the laser pulse.
Significantly, we observe that the spin density and spin fluxes in the domain wall depend on the
distance of the excitation region from the domain wall, and therefore extend our study to investigate how the spin-transfer torque acting on the domain wall can be manipulated 
changing the position of the excitation region along the sample.
Consequently, we explore domain wall dynamics excited by the spin-transfer torque
of superdiffusive hot electrons.

To inspect the generation of the spin-transfer torque by the laser pulse, 
we assume that the domain wall is located in the middle of the sample, $z = 0$.
The laser pulse excites hot electrons in the excitation region. Part of the hot electrons
pass the domain wall and the spin flow locally generates spin-transfer torque due to magnetization variation.
The spin-transfer torque is proportional to the local transverse spin current as given by Eq.~(\ref{Eq:def_tor}).

%%%%%%%%%%%%%%%%%%%%%%%%%%%%%
\begin{figure}[!htp]
 \centering
 \includegraphics[width=.9\columnwidth]{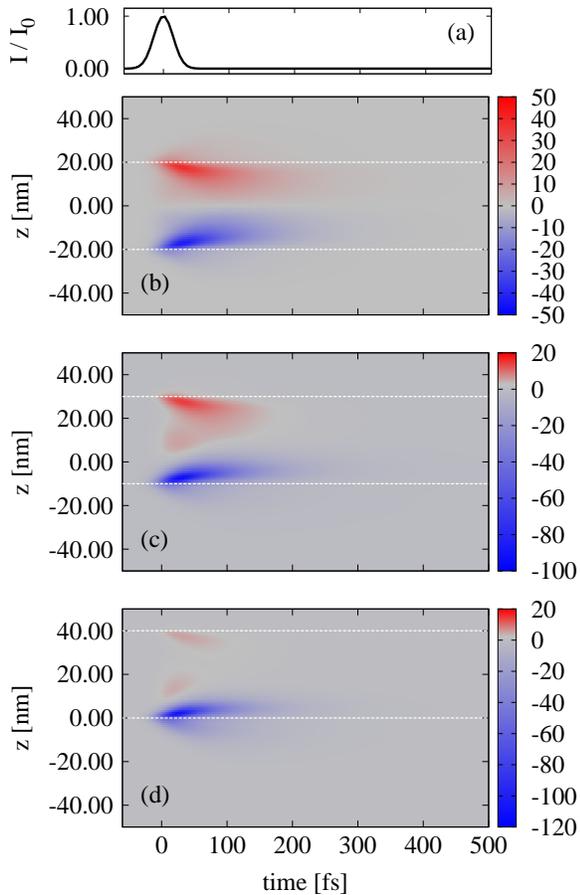}
 \caption{(Color online) Spatial and time dependence of the transverse spin current in the neighborhood of a domain wall 
          in the units of $[\hbar/(2 e)]\, {\rm fs}^{-1}$. The domain wall center is located in $z = 0$ and its width is $\Delta = 10\, {\rm nm}$.
          The excitation region is restricted to the area between the white dashed lines.
          (a) Time variation of Gaussian laser pulse intensity with peak at time $t = 0$ and FWHM $35\, {\rm fs}$.
          (b) The excitation region is symmetric with respect to the position of the domain wall center.
          (c) and (d) The excitation region is asymmetric with respect to the position of the domain wall center.
          In case (d) the borderline of the excitation region passes through the domain wall center.}
 \label{Fig:torques}
\end{figure}
%%%%%%%%%%%%%%%%%%%%%%%%%%%%%

Fig.~\ref{Fig:torques} shows the space-time maps of the spin-transfer torque in the neighborhood of the domain wall.
The DW width is $\Delta = 10\, {\rm nm}$ and the width of the excitation region is assumed to be $l_{\rm ex} = 40\, {\rm nm}$.
The temporal profile of the laser beam intensity used in the calculations is shown in Fig.~\ref{Fig:torques}(a).
Fig.~\ref{Fig:torques}(b) depicts a situation when the excitation region is symmetric with respect to the domain wall center.
Importantly, the spin-transfer torque in the center of the domain wall remains zero all the time.
Conversely, on both sides of the domain wall the spin torque is nonzero and changes in time. 
This torque is nonuniform in space and varies in time. 
Its variation is caused by both position dependence of the magnetization gradient across the domain wall
as well as spin relaxation.
Due to the symmetry of the system, the spatial dependence of the spin-transfer torque
remains antisymmetric.
The spin-transfer torque decreases in time; the total time in which the spin torque is acting on the magnetic moments is
about $\sim 500\, {\rm fs}$.

The situation is different when the center of the excitation region is shifted from the DW center.
This is shown in Figs.~\ref{Fig:torques}(c) and (d), which depict the transverse spin current
when the center of the excitation region is shifted from the domain wall center by $10$ and $20\, {\rm nm}$, respectively.
In both cases we observe asymmetry of the spin currents on the right and left hand side of the domain wall.
As a result, the spin torques acting in the domain wall become strongly asymmetric.
Fig.~\ref{Fig:torques}(d) shows a specific case, where one of the borderlines of the excitation region is located at the DW center.
Thus, electrons are excited by the laser pulse just in one half of the domain wall.
The dominant spin-transfer torque is generated in the vicinity of the domain wall center at the borderline of the excitation region.

Summarizing, Figure \ref{Fig:torques} illustrates that a symmetric DW excitation leads to zero total spin-transfer torque and no domain wall motion. Additionally, as the center of the excitation region departs from the DW center, the asymmetry of the spin torque increases, and as a result magnetization dynamics is expected.
This magnetization dynamics can possibly lead either to a deformation of the domain wall structure or to a domain wall motion.
A combination of both effects is also possible.

In the previous section we have shown that laser-induced spin transport between the domains can reduce the magnetization by about $20\%$.
Such a reduction of magnetization can affect both the local effective fields acting on the localized magnetic moment and the spin- transfer torque.
The response of the magnetization dynamics to the variation of an effective magnetic field in ferromagnets will happen on a nanosecond scale, which
is definitely longer than the domain wall dynamics induced by superdiffusive spin-transfer torque.
For this reason we do not assume a magnetization reduction in the effective magnetic field, but this effect is included in the spin-torque term. 
Similarly, in our simulations we disregard effects connected with temperature gradients or change of the magnetic anisotropy.

\subsection{Domain wall dynamics}

By using the LLG  model (Eq.\ (\ref{Eq:LLG})) we study how the spin-transfer torque influences the magnetization dynamics.
We start our simulations from a static configuration with a head-to-head magnetic domain wall located
in the center of the sample. Magnetic moments completely lay in the plane of the sample.
The magnetization dynamics starts with the $35$\,fs laser pulse leading to a time-dependent spin torque acting on the localized magnetic moments in the chain.

In our simulations we have assumed an
equilibrium saturated magnetization $M_0 = 1.7 \times 10^6\, {\rm A/m}$,
exchange stiffness $\Aex = 2 \times 10^{-11}\, {\rm J}/{\rm m}^3$, 
and no applied magnetic field. The width of the magnetic domain wall
will be modified by the uniaxial magnetic anisotropy, $\Ku$, which
obeys Eq.\ (\ref{Eq:DW_width}).
The distance between the localized magnetic moments is in agreement with the discretization 
in the spin transport calculations, i.e.\ $a = 1\, {\rm nm}$.
Finally, the Gilbert damping parameter $\alpha = 10^{-3}$ has been assumed.

%Importantly, in experimental conditions change of magnetization can influence
%the spin transport and thus spin transfer torque can depend on the magnetization dynamics.
In our simulations we assume that the magnetization dynamics will be small enough and
will not substantially influence the flow of electrons.
Therefore, we can separate the simulation of magnetization dynamics from the ones of superdiffusive spin-dependent transport.

Initially, our simulations reveal that the laser pulse primarily causes a shift of the center of the magnetic domain wall
without any substantial modification of the DW profile or magnetization tilting.
This simplifies the description of the DW dynamics to the time dependence of the DW center position.
%%%%%%%%%%%%%%%%%%%
\begin{figure}
 \centering
 \includegraphics[width=.8\columnwidth]{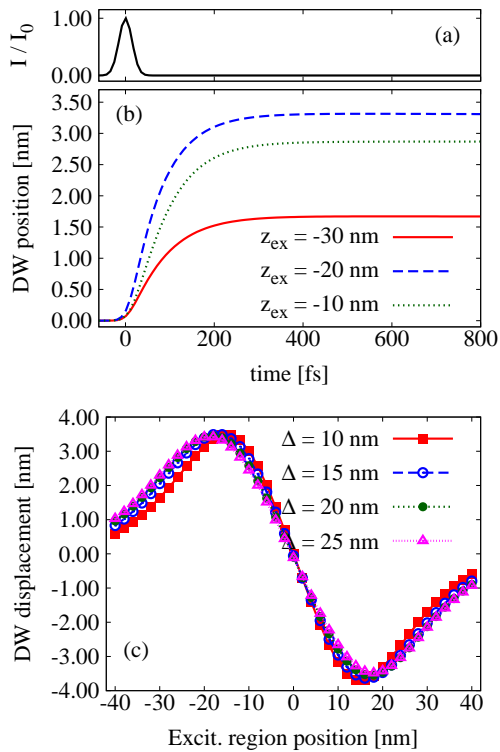}
 \caption{(Color online) Laser-induced dynamics of the domain wall.
          (a) Time variation of Gaussian laser pulse intensity with peak at time $t = 0$ and FWHM $35\, {\rm fs}$.
          (b) Time dependence of the domain wall position after the laser pulse for DW width $\Delta = 10\, {\rm nm}$ 
          calculated for different positions of the excitation region.
          (c) DW displacement taken $1\, {\rm ps}$ after the DW motion for different DW widths 
          as a function of the excitation region position.}
 \label{Fig:DW_dynam}
\end{figure}
%%%%%%%%%%%%%%%%%%%
Therefore, in Fig.~\ref{Fig:DW_dynam}(b) we show the time evolution of the DW center after the laser pulse is applied.
Different curves correspond to different positions of the excitation region.
The position of the excitation region, $\zex$, is given by its center with respect to the initial position of the DW center.
In all our calculations the length of the excitation region is $l_{\rm ex} = 40\, {\rm nm}$.
After the pulse is applied, the domain wall starts to move. It stops after about $500\, {\rm fs}$, which
corresponds to the time when the spin fluxes diminish.
Importantly, the DW displacement caused by one laser pulse strongly depends on the position of the excitation region.
This means that also the DW velocity depends on $\zex$.

Fig.~\ref{Fig:DW_dynam}(c) shows how the DW displacement depends on the position of the excitation region.
To this end we show the DW position $1\, {\rm ps}$ after the laser pulse maximum when its dynamics completely stopped.
The dependence is shown for various DW widths.
The plot shows nonmonotonic dependence of DW displacement on $\zex$ revealing a number of properties.
We observe that, for all values of $\Delta$, DW displacement remains zero for a symmetric electron excitation when $\zex = 0$.
This is caused by zero total spin-transfer torque produced in the symmetric case, as explained by Fig.~\ref{Fig:torques}(b).
Besides, as the center of the excitation region departs from the DW center, the absolute value of the DW displacement increases.
The increasing trend persist up to a certain maximum value. This value is located for all studied DW widths at $\zex \simeq \pm 20\, {\rm nm}$, which
correspond to the cases when one of the borderlines of the excitation region
is located at the initial position of the DW center $z = 0$.
Small deviation of the extremes in the dependence shown in Fig.~\ref{Fig:DW_dynam}(c)
from $\zex = \pm 20\, {\rm nm}$
might be caused by minor modification of the DW structure during the dynamics induced by
nonhomogeneous spin currents.
In this case, the asymmetry of right and left electron fluxes is maximal, 
which also maximizes the 
total spin-transfer torque acting on the DW, as shown in Fig.~\ref{Fig:torques}(d).
We also find that, when the distance of the excitation region becomes larger than $\zex = l_{\rm ex}/2$, 
the DW displacement decreases. This reduction of the DW displacement is related
to the spin relaxation of the itinerant electrons which have to go through longer distances
to reach the domain wall and contribute to the spin-transfer torque.
Finally, the dependence of DW displacement on $\zex$ is an odd function, which
depends on the direction of dominant flux of electrons passing the domain wall.
If the excitation region position is shifted to the left (right) from the initial position of
the DW center, the dominant electrons flux contribution to the spin-transfer torque will be oriented to the right (left).
Thus, the domain wall moves to the right (left) in agreement with the incident spin current.

\section{Discussion}
\label{Sec:Discussion}

In Sec.~\ref{Sec:NarrowDW} we have studied a simplified model of a narrow magnetic domain wall.
We find that our theoretical calculations are comparable to experimental observations and Monte Carlo simulations~\cite{Pfau2012}
showing a broadening of magnetic domain walls on a femtosecond timescale.
This model was, however, restricted to collinear magnetic moments and therefore spin-transfer torques could not be analyzed. 

As a next step, in Sec.~\ref{Sec:WideDW} we have focused on wider domain walls taking into account the details
of the domain wall profile. In this case we have found an enhancement of demagnetization
in the vicinity of the domain wall.
Similarly, as in the case of a narrow domain wall, the demagnetization effect
is caused by the transfer of superdiffusive hot electrons between magnetic domains of opposite magnetization direction.
In real 3-dimensional samples, this effect should appear on top of other processes leading to
ultrafast demagnetization as a modification of the domain wall profile,
which appears on a time scale of a few hundreds of femtoseconds.
The second effect studied here is the generation of a spin-transfer torque due to a 
femtosecond laser pulse focused on a narrow part of the sample close to the domain wall. 
We could show that when the area excited by the laser pulse is asymmetric with respect to the domain wall center,
the interaction of the hot electrons with the magnetization gradient
can create a directed spin-transfer torque strong enough to induce domain wall motion.
Due to the transient nature of the superdiffusive spin currents the spin-transfer torque acts only within less than $1\, {\rm ps}$.
The resulting DW displacement is not large, $\sim 3.5\, {\rm nm}$ (Fig.~\ref{Fig:DW_dynam}), yet it takes place very fast, which hence implies  a significant rapidly changing out-of-equilibirum DW velocity, 
whose average value over first $250\, {\rm fs}$ is 
$\sim 1.4 \times 10^4\, {\rm m}/{\rm s}$ (for $z_{\rm ex} = 30\, {\rm nm}$). 
This value exceeds substantially currently known DW velocities.  For example, high DW velocities ($\sim 750\, {\rm m}/{\rm s}$) have so far only been reported for synthetic antiferromagnets~\cite{Yang2015}.
We further note that current or magnetic field induced DW motions have been studied under quasi-equilibrium conditions, under which the DW velocity is limited, e.g.~ by the Walker breakdown limit~\cite{SchryerWalker1974}. Beating this limit has been an issue for some time~\cite{Yan2010,Selzer2016}; using out-of-equilibrium spin-current pulses could offer a route to overcome this limit.

Despite the qualitative agreement with experimental results, it is at this point relevant to mention the assumptions considered in our model and the constrains they impose on the magnetization dynamics due to spin transfer.
Currents of superdiffusive hot electrons persist in the sample for about $500\, {\rm fs}$ \cite{Battiato2012}.
This is the time scale of the demagnetization process and spin-transfer torque action.
The question is how these effects can finally influence the resulting magnetization dynamics.
First, the partial ultrafast demagnetization, which happens on the same timescale as the
spin-transfer torque, influences just the magnitude of the magnetization. 
In our calculations e.g.\ for a $10\, {\rm nm}$ domain wall width this effect is about $15\,\%$
and it strongly decreases with the domain wall thickness.
Importantly, the direction of magnetization remains unchanged.
Note that for a reduced magnetization the torque actually causes a bigger change in the magnetization.
The inhomogeneous variation of magnetization length, $M(z)$, can induce changes in the local effective magnetic field and, consequently, magnetization dynamics can occur.
However, this magnetization dynamics in the local magnetic field has typically a timescale of a few nanoseconds~\cite{Berkov_Miltat_2008}.
Therefore, it is out of the scope of this work. 
Second, in our modeling of the magnetization dynamics we assumed that the domain wall motion
does not influence the spin transport.
This assumption is maybe more problematic since both domain wall motion and spin-transfer torque are mutually coupled and share the same time scale.
Our argument is based on the length scale and on the smoothness of the domain wall profile. 
Namely, in the domain wall the magnetization is varying slowly. When the domain wall shifts by $\sim 3\, {\rm nm}$,
the generated spin torque is not substantially influenced by a minor change of local magnetic direction.
More important for the modeling is the effect of relaxation of the hot electrons during the transport.
Although the former effect is not taken into account in our simulations, the latter one
is fully incorporated.

In addition, it deserves to be mentioned that a femtosecond laser pulse might be the source of other effects leading to domain wall motion, which are not included in our simulations, such as the entropic torque~\cite{Schlickeiser2014:PRL,Wang2014:PRB,KimSeKwon2015:PRB}, 
which forces the domain wall to move towards the laser spot, and the magnonic torque~\cite{YanP2011:PRL,KimSeKwon2015:PRB,HIN-11} of thermally induced magnons that can influence the domain wall motion.
Although the former is relatively strong, it leads to magnetization dynamics on a nanosecond timescale and
domain wall velocities reach values of $\sim 10^2\, {\rm m}/{\rm s}$~\cite{Moretti2017:PRB}.
Importantly, the direction of the entropic torque is opposite to the one induced by superdiffusive spin-dependent transport. On the other hand, the magnonic torque is relatively weak and leads to domain wall velocities of the order of only $10\, {\rm m}/{\rm s}$~\cite{Moretti2017:PRB}.

Lastly, a further aspect that becomes important for domain wall motion once the fast spin-transfer torque has ceased, is the domain wall inertia (see, e.g.~\cite{Thomas2010,Janda2017})
which we have not considered here. While the superdiffusive spin currents acts only within one ps, these could provide a stimulus that enables depinning of domain walls \cite{Sandig2016} and initiate inertial domain wall motion.

Finally, we have studied a domain wall in a material with in-plane uniaxial magnetocrystalline anisotropy, where the magnetization direction varies smoothly on a long length scale.
Nevertheless, our model allows also to study sharp magnetic domain walls of arbitrary profile, which can be observed
in multilayers with strong perpendicular magnetic anisotropy~\cite{Pfau2012,r_12_Vodung_UltraDemagMultilay_SpinTransp,Moisan2014,Quessab2018:PRB} 
or in geometrically constrained domain walls~\cite{Bruno1999:PRL}.

\section{Conclusions}

In summary, we have formulated a model describing 1-dimensional laser--generated transport of hot electrons
through a magnetic domain wall and the spin dynamics it induces. Our study demonstrates both the
contribution of the spin transport between magnetic domains of opposite spin direction
to the ultrafast demagnetization as well as the possibility of spin torque generation.
We have shown that when the laser beam is focused on a restricted area of the sample, 
it can create an imbalance between right and left flowing fluxes of hot electrons
flowing through the domain wall. This leads to the nonzero total spin-transfer torque that can
induce a shift of the domain wall by few nanometers in about $500\, {\rm fs}$.
This new mechanism of domain wall motion creates a relatively small shift compared to other laser-induced mechanisms, 
like entropic~\cite{Schlickeiser2014:PRL,Wang2014:PRB,KimSeKwon2015:PRB} or magnonic~\cite{Moretti2017:PRB} torques due to thermal magnons,
when one looks at the situation tens of picoseconds after the pulse. 
However, a definite advantage is that it can be controlled by femtosecond laser pulses, and within the time window of less than a ps it provides very high DW velocities of the order of $10^4\, {\rm m/s}$ in a ferromagnetic system,
which exceeds considerably the values of velocities of current-induced domain wall motion~\cite{LiZhang2004:PRB,Gonzales2012:PRL,LeeJY2007:APL}.

\section*{Acknowledgement}

This work was supported by the European Regional Development Fund 
in the IT4Innovations national supercomputing center - path to exascale project 
(project number CZ.02.1.01/0.0/0.0/16\_013/0001791) within the Operational Programme Research, Development and Education,
 by the Czech Science Foundation (grant number 18-07172S), by the Swedish Research Council (VR), and the K.\ and A.\ Wallenberg Foundation (grant No.\ 2015.060).
The authors furthermore thank the Ministry of Education, Youth and Sports for the Large Infrastructures for Research, Experimental Development and Innovations 
project ``IT4Innovations National Supercomputing Center LM2015070", as well as the Deutsche Forschungsmeinschaft for financial support via RI 2891/1-1, and the Swedish National Infrastructure for Computing (SNIC). We thank J{\'e}r{\^o}me Hurst for valuable discussions.

\bibliographystyle{apsrev4-1}
%%% \bibliography{bib_neu}
%

\end{document}